\documentclass[a4paper,11pt]{article} 
\usepackage[dvips]{graphics} 
\usepackage[british]{babel} 
\usepackage{amssymb,amsmath} 
\usepackage[ansinew]{inputenc} 
\usepackage{epsfig} 
\begin{document} 
\begin{center} 
\Large\textbf{{In defence of non-ontic accounts of quantum states}}\vspace{0.5cm}\\ 
\normalsize Simon Friederich\vspace{0.5cm}\\ 
\texttt{email@simonfriederich.eu}\vspace{0.2cm}\\ 
Philosophisches Seminar, Universit\"at G\"ottingen, Humboldtallee 19, D-37073 G\"ottingen, Germany 
\end{center}

\small{ 
\noindent The paper discusses objections against non-hidden variable versions of the epistemic conception of quantum states---the view that quantum states do not describe the properties of quantum systems but reflect, in some way to be specified, the epistemic conditions of agents assigning them. In the first half of the paper, the main motivation for the epistemic conception of quantum states is sketched, and a version of it is outlined, which combines ideas from an earlier study of it \cite{Friederich} with elements of Richard Healey's recent pragmatist interpretation of quantum theory \cite{Healey}. In the second half, various objections against epistemic accounts of quantum states are discussed in detail, which are based on criticisms found in the literature. Possible answers by the version outlined here are compared with answers from the quantum Bayesian point of view, which is at present the most discussed version of the epistemic conception of quantum states.
}\vspace{0.3cm}\\ 

\noindent Keywords: quantum states, quantum probabilities, anthropocentric notions, micro/macro divide, explanation and prediction\vspace{0.3cm}\\ 

\section{Introduction}

The measurement problem and the problem of quantum ``non-locality'', that is, the claimed tension between quantum theory and relativity theory, are widely regarded as the most outstanding difficulties in the foundations of quantum mechanics. Possible ways to react to these problems (or ``paradoxes'') range from changing the dynamics (as in GRW theory) to adding determinate particle and field configurations (as in pilot wave approaches) to adopting a non-standard picture of our world according to which our universe (or our mind) constantly splits into an immense number of branches (as in variants of the Everett interpretation). These are attempts to \textit{solve} the paradoxes, either by altering the formalism of the theory or by radically altering our picture of the world so that at least one of the assumptions necessary to derive the paradoxes no longer holds.

The present paper investigates accounts of quantum theory which approach the paradoxes from an entirely different perspective. Their main ambition is to \textit{dissolve} the paradoxes by proposing a perspective on the linguistic roles of the constituents of the quantum theoretical formalism according to which at least one of the assumptions necessary to derive the paradoxes is not wrong (in which case one might speak of a \textit{solution} rather than a dissolution) but conceptually ill-formed. The motivation underlying these approaches is ``therapeutic'' inasmuch as they aim at ``curing'' us from what they see as unfounded worries about foundational issues based on conceptual misunderstandings. More specifically, the accounts to be discussed are grounded in the hope that a \textit{non-ontic} reading of quantum states---a reading which construes quantum states as having a non-descriptive linguistic role, which differs from that of representing reality---may hold the key to conceive of the theory in such a way that the paradoxes do not arise in the first place.

The defining characteristic of non-ontic readings of quantum states is that they reject the notion of a \textit{true} quantum state of a quantum system---a quantum state it \textit{is in}.\footnote{According to the terminological conventions employed in this paper, an account of quantum states qualifies as ``non-ontic'' if and only if it denies that for every quantum system there exists exactly one (true) quantum state it is in. Accounts which accept this assumption are referred to as ``ontic''. Thus, accounts need not attribute any metaphysically ambitious kinds of reality to quantum states (as some versions of pilot wave theory may not) in order to count as ontic in the sense used here.} Non-ontic readings of quantum states mostly embrace some version of the \textit{epistemic} conception of quantum states---the view that quantum states somehow reflect the state-assigning agents' epistemic relations to the systems states are assigned to\footnote{There are versions of the epistemic conception of states which add additional (ontic) variable configurations to the standard formalism of quantum theory. Important contributions to this type of approach include the construction of an explicit toy model based on the epistemic conception of quantum states by Spekkens (see \cite{Spekkens}), which reproduces many signature qualitative features of quantum theory. More recently, a theorem due to Pusey, Barrett and Rudolph (see \cite{PBR}) has stimulated a lot of interest and research activity, which establishes that an important class of hidden-variable models based on the epistemic conception yields predictions which are compatible with those of standard quantum theory. Hidden-variable models combined with epistemic accounts of quantum states do not conform to the therapeutic ambitions outlined before and are therefore not further discussed in what follows, but their investigation represents no doubt an intriguing and flourishing field of research.}---and the present paper focuses on these. A defining criterion of what counts as a version of the epistemic conception of states is to allow that agents which are differently situated with respect to one and the same quantum system and have different epistemic relations to it may (and perhaps even should) legitimately assign different quantum states to it. As formulated by Rudolf Peierls, one of its key proponents, the quantum states assigned to the same system by different agents ``may differ as the nature and amount of knowledge may differ.''\footnote{See \cite{Peierls} p.\ 19.}

Accounts that are based on the epistemic conception of states must be non-ontic in that they cannot acknowledge any such thing as an agent-independent ``true'' quantum state of a quantum system---a state it ``is in''---, for if such a state exists, assigning precisely that state is the one and only correct way of assigning a quantum state and assigning any other state counts as wrong. Any account that accepts the notion of a ``quantum state a quantum system is in''---and does not interpret it as a sometimes harmless yet strictly speaking misleading \textit{fa\c{c}on de parler}---does not qualify as an epistemic account of quantum states in the sense of the present paper.

Just like any other take on the foundations of quantum theory, the epistemic conception of states has been heavily criticised on a number of grounds. The main aim of the present paper is to consider the most important objections brought forward together with possible rejoinders. Most of the objections considered were originally raised as criticisms of \textit{quantum Bayesianism}, a remarkably radical epistemic account of quantum states the core idea of which is to interpret quantum probabilities as subjective degrees of belief in accordance with the subjective Bayesian account of probability. One of the goals of the present paper is to explore quantum Bayesianism's resources for answering these objections.

In addition to quantum Bayesianism, the paper expounds and develops another version of the epistemic conception of quantum states, introduced in the following sections as the ``Rule Perspective'' (for reasons to be given). It is based on a combination of ideas developed in an earlier study of the epistemic conception of states (see \cite{Friederich}) with elements drawn from Richard Healey's recent pragmatist interpretation of quantum theory (see \cite{Healey}).\footnote{\label{fussnote}%For studies elaborating on the epistemic conception of states and views in a similar spirit see \cite{FuchsPeres}, \cite{Merminnew}, \cite{Caves2002a}, \cite{Caves2002b}, \cite{Fuchs}, \cite{Pitowsky}, \cite{Appleby}, \cite{Bub}, \cite{Caves2007}, \cite{FuchsSchack1}, \cite{BubPit}, \cite{Barnum}, \cite{FuchsSchack2}, \cite{Friederich}.

Healey's pragmatist interpretation is closely related to the epistemic conception of quantum states in many respects, but Healey himself does not regard his own view as a version of the epistemic conception (see \cite{Healey} p.\ 16). The similarity in spirit becomes clear, for instance, in Healey's take on measurement collapse ``as a way of updating [an agent's] authoritative source of advice'', in complete agreement with the perspective on collapse suggested by epistemic accounts of quantum states, see Section 2.1.} The main result of the discussion of objections in later sections of this paper is that the Rule Perspective fares much better in answering them than quantum Bayesianism. This is not meant to suggest that the Rule Perspective is superior to all other rival interpretive takes on quantum theory or that it even constitutes the uniquely correct account of quantum foundations. The more modest aim of this paper is to argue that the idea of dissolving the paradoxes without introducing any hidden variables, additional dynamics and without accepting Everettian branching can be spelled out and defended in a coherent way. Arguably, this suffices to make accounts that are based on it into serious contenders among interpretations of quantum theory.

The structure of the remaining sections of this paper is as follows: Section 2 sketches in which way the epistemic conception of quantum states dissolves the notorious paradoxes of measurement and non-locality by undermining their conceptual presuppositions. Section 3 reviews quantum Bayesianism and introduces the Rule Perspective. Sections 4, 5, and 6 are the core sections of the paper. They consider objections against the epistemic conception of quantum states (found in the literature or inspired by it) and develop and discuss possible answers from the points of view of quantum Bayesianism and the Rule Perspective. The paper closes in Section 7 with some remarks on the shared ambition of the Rule Perspective and the Everett Interpretation to understand quantum theory without supplementing its standard formalism in any way.

\section{Dissolving the paradoxes}
The present section gives an outline of how the epistemic conception of quantum states dissolves the measurement problem and the problem of quantum ``non-locality'', that is, the claimed tension between quantum theory and relativity theory. I consider the measurement problem first.

\subsection{The measurement problem}
There are various different formulations of the measurement problem in the literature, sometimes even conceived of as different measurement problems.\footnote{See \cite{Wallace} for a detailed ``modern'' take on the measurement problem and an overview of proposed solutions.} Here I focus on an exposition due to \cite{maudlin95} (who alone gives three different formulations), to illustrate how the epistemic conception of quantum states dissolves it. In Maudlin's formulation, the problem arises from the incompatibility of the following three assumptions:
\begin{quote}
1.A The wave-function of a system is complete, i.e. the wave-function specifies (directly or indirectly) all of the physical properties of a system.\\
1.B The wave-function always evolves in accord with a linear dynamical equation (e.g. the Schr\"odinger equation).\\
1.C Measurements of, e.g., the spin of an electron always (or at least usually) have determinate outcomes, i.e., at the end of the measurement the measuring device is either in a state which indicates spin up (and not down) or spin down (and not up). (\cite{maudlin95} p.\ 7.)
\end{quote}
To see that these assumptions are incompatible (though not strictly logically inconsistent, see \cite{maudlin95} p.\ 10), one has to consider a situation in which the system being measured is not in an eigenstate of the measured observable. In that case, according to 1.B, the state of the combined system consisting of the measured system together with the measuring apparatus evolves into a superposition of eigenstates of the measured and the ``pointer'' observable, whose different possible values correspond to macroscopically different configurations of the ``pointer'' (or display) of the apparatus. This state doesn't single out or prefer by itself any of the possible values of the measured observable of the measured system nor of the pointer observable pertaining to the apparatus. Nevertheless, according to 1.A, it provides a complete description of the combined system, including both the measured system and the apparatus. On the basis of this assumption we have to conclude that none of the possible values of the measured and the pointer observables is actually realised (or all at once, as Everettians would conclude). Assumption 1.C, however, requires that at the end of the measurement process the value of the pointer observable is determinate. Consequently, the three assumptions, taken together, are incompatible and at least one of them has to be dropped. A necessary condition for an interpretation of quantum theory to count as a candidate \textit{solution} to the measurement problem is that it declares either 1.A or 1.B or 1.C to be wrong (or that it finds a loophole in the reasoning leading to their claimed incompatibility, which seems very difficult). Solutions to the measurement problem contrast with \textit{dissolutions}. Dissolutions also reject at least one of the assumptions 1.A, 1.B or 1.C---but for being senseless rather than for being wrong or, to put it more diplomatically, for being based on mistaken conceptual presuppositions rather than for being mistaken.

Accounts that are based on the epistemic conception of states dissolve the measurement problem in precisely that sense. They do not acknowledge the notion of a quantum state a quantum system ``is in'' and reject the view that it is part of the linguistic role of quantum states to specify ``physical properties of a system'' (Maudlin's assumption 1.A) in the first place. Whereas solutions to the measurement problem which reject assumption 1.A as \textit{wrong} hold that quantum states are \textit{incomplete} specifications of the physical properties of quantum systems (pilot wave approaches, for example, inasmuch as according to them the quantum state specifies \textit{some} features of physical reality, though not all), the epistemic conception of states denies that quantum states are appropriately regarded as specifications of physical properties in the first place. Assumption 1.C (outcome determinateness) is left untouched, and the ramifications for assumption 1.B are essentially the same as for 1.A since the notion of ``\textit{the} wave-function'' of a system is rejected, as it presupposes that of a quantum state a quantum system is in. Agents who are competent in applying quantum theory do of course \textit{assign} quantum states to quantum systems and they \textit{make them} undergo unitary time-evolution according to the Schr\"odinger equation, but acknowledging this is very different from endorsing an interpretation of unitary time-evolution as corresponding to the evolution of physical properties of the system itself.

In the application of quantum theory the measurement problem does not arise as a practical difficulty, as it is circumvented by what is widely seen as an act of brute force, namely, by invoking the notorious von Neumann measurement collapse of the wave-function. From the perspective of ontic accounts of quantum states collapse is disconcerting in that it remains completely dubious and unclear under which conditions it occurs. From the perspective of the epistemic conception of states, in contrast, measurement collapse merely reflects a sudden and discontinuous change in the epistemic situation of the state-assigning agent, not a discontinuity in the time-evolution of the system itself, so the question of under which conditions it occurs becomes meaningless. A question that does make sense from the point of view of the epistemic conception of quantum states is under which conditions an agent assigning a quantum state \textit{should apply} the collapse postulate. This question will be discussed in more detail in connection with the Rule Perspective in Section 3.2. Objections against the employment of ``anthropocentric notions'' (such as ``measurement'', ``information'', ``apparatus'') in this take on collapse are discussed in Section 5.

The epistemic conception of states undermines the measurement problem in the form just discussed and offers an elegant justification of the otherwise mysterious collapse, but it cannot guarantee that the problem will not arise at a different stage in a different form. One objection critics are likely to raise, for instance, is that rejecting conceptual presuppositions of the measurement problem hardly suffices to account for why measurements have determinate outcomes in the first place. No-go theorems on assigning determinate values to observables (the Kochen-Specker theorem and its relatives in particular) make it unattractive to assume that observables have determinate values at all times, so the question arises why precisely those observables which we measure never fail to exhibit determinate values at the end of a measurement process. Even though the epistemic conception of states dissolves the measurement problem in the sense of a sharp antinomy of conflicting assumptions, it does not by itself account for why measurements always seem to result in determinate outcomes. Different epistemic accounts of quantum states respond differently to this challenge, and I postpone the discussion of possible answers to a later stage (see Section 5.2).

\subsection{The problem of ``non-locality''}
The problem of whether quantum theory is compatible with the principles of relativity theory arises perhaps most blatantly in the challenge of reconciling the time-evolution of quantum states including collapse in whatever suitable form with the requirement of Lorentz invariance as inferred from relativity theory. To see this problem in an example, consider a two-particle system in an EPR-Bohm setup where two spin-$1/2$ systems $A$ and $B$ are prepared in such a way that those knowing about the preparation procedure assign an entangled state, for instance, the state $\frac{1}{2} (|+\rangle_A|-\rangle_B-|-\rangle_A|+\rangle_B)$ for their combined spin degrees of freedom. Assume further that the two systems $A$ and $B$ have been brought far apart and an agent Alice, located at the first system $A$, measures its spin in a certain direction. Having registered the result and having applied the projection postulate, she assigns two no longer entangled states to $A$ and $B$, which depend on the choice of observable measured and on the measured result. Another agent, Bob, located at the second system $B$, may also perform a spin measurement (in the same or in a different direction of spin) and proceed to assign a pair of no longer entangled states to the two systems in an analogous way. Now the intriguing challenge concerning the compatibility of quantum theory and relativity theory is to specify at which time which system is in which state and to do so in a Lorentz invariant manner. The difficulty is especially blatant for cases where the measurements carried out by Alice and Bob occur in spacelike separated regions, perhaps even in such a way that each of them precedes the other in its own rest frame (see \cite{Zbinden}). In that case there is clearly no non-arbitrary answer to the question as to which measurement occurs first and triggers the abrupt change of state of the other. There exist attempts of overcoming this problem without completely abandoning either collapse or the notion of a quantum state a quantum system is in by making the time-evolution of quantum states dependent on foliations of space-time into sets of parallel hyperplanes, but there does not seem to be general agreement as to whether this programme succeeds, and the approach remains controversial.\footnote{See \cite{Fleming} for a hyperplane-dependent formulation of state reduction and \cite{Myrvold} for an instructive defence of that approach. For criticism see \cite{Maudlin}, which, based on the presupposition that the ontic conception of states is correct, comes to a very negative verdict on whether relativity theory and quantum theory in its present form can be consistently combined at all.} 

The epistemic conception of quantum states undermines the conceptual presuppositions of the reasoning leading to this problem by rejecting the notion of a quantum state a quantum system is in and by interpreting the assignment of different quantum states to the two systems by different agents as legitimate and very natural: Alice knows about the preparation procedure for the combined two-particle system, and when she registers the result pertaining to her own system this affects her epistemic condition with respect to the second. The state she assigns to it reflects her epistemic relation to it, and there is no need to assume that her measurement of her own system has a physical influence on the second. The same considerations apply for Bob. Predictions for the results of measurements derived on the basis of entangled states may still be baffling and unexpected, but the dynamics of quantum states do not give rise to any incompatibility between the principles of relativity theory and those of quantum theory as construed by the epistemic conception of states. The paradox of quantum non-locality (inasmuch as it resides in the challenge of reconciling the time-evolution of quantum states with the principles of relativity theory) is \textit{dissolved} by rejecting its conceptual presupposition, namely, the ontic conception of states and the interpretation of the time-evolution of quantum states as a physical process.

In analogy to the case of the measurement problem, the dissolution of the problem of ``non-locality'' just outlined is no demonstration that there are no further problems linked to reconciling quantum theory with relativity theory. Several authors claim that there are indeed serious problems, for example that quantum theory seems to violate Bell's principle of \textit{local causality}\footnote{See \cite{Bell90} for a helpful introduction to that concept and \cite{Seevinck} for a recent extension and sharpening of Bell's claims and results.}, which, as it may seem, can only be accounted for by assuming superluminal causation. Healey argues against this claim that Bell's concept of local causality is simply inapplicable to quantum theory and that ``non-local'' correlations obtained from quantum theory are perfectly compatible with relativity theory.\footnote{See \cite{Healey2}.} So, the suggested violation of local causality in quantum theory and the associated claim concerning the incompatibility between quantum theory and relativity theory remain controversial.

Another route for suggesting that there is a severe tension between quantum theory and relativity theory is to argue that quantum theoretical ``non-local'' correlations ground counterfactual claims which are only true in the presence of superluminal causation.\footnote{See, for instance, \cite{Butterfield} and Chapter 5 of \cite{Maudlin} for such arguments.} Such arguments do not necessarily presuppose any specific view of quantum states, and, inasmuch as they don't, other moves are required to address them than adopting a non-ontic account of quantum states. The most promising strategies for responding are either to question that quantum theory really licences these counterfactuals at issue or to deny that these counterfactualy are concerned with causal connections in the first place.\footnote{See \cite{Healey2}, Section 7, for a detailed defence of the view that the connections reported in counterfactual statements regarding ``non-local'' correlations are not causal and \cite{Stairs} for considerations against the view that ``EPR-type cases sanction counterfactual inferences'' at all. (\cite{Stairs} p.\ 165.)} Fortunately, it is largely irrelevant for the purposes of the present paper whether or not such counterfactuals deserve to be called ``causal'' as long as no incompatibility between quantum theory and relativity theory arises in a more straightforward way. The epistemic conception of states may not completely dispel the felt tension between quantum and relativity theory, but it arguably removes any clear-cut inconsistency between the two.

\section{Two epistemic accounts of quantum states} 
Various different epistemic accounts of quantum states have been proposed, and it is not my aim to provide a systematic overview of them here. Rather, I shall focus only on two (very different) epistemic accounts of quantum states and discuss their respective problems and advantages in more detail. The first account is quantum Bayesianism, a position developed by C. M. Caves, C. A. Fuchs and R. Schack, supported in some of its central tenets by H. N. Barnum, M. D. Appleby (and perhaps other authors).\footnote{See \cite{Caves2002a}, \cite{Caves2002b}, \cite{Fuchs}, \cite{Appleby}, \cite{Caves2007},\cite{FuchsSchack1}, \cite{FuchsSchack2}, \cite{Barnum} for contributions and (essentially) approving comments to the quantum Bayesian programme and \cite{Timpson} for a highly useful review and critique.} Its core idea is to interpret quantum probabilities as subjective degrees of belief in accordance with the subjective Bayesian take on probability. An exposition of the central claims and results of quantum Bayesianism is given in Section 3.1. The second account to be discussed, the \textit{Rule Perspective}, combines ideas developed in \cite{Friederich} with ideas drawn from Richard Healey's recently proposed pragmatist interpretation of quantum theory (see \cite{Healey}). Section 3.2.1 introduces it as an account of the rules governing the assignment of quantum states as \textit{constitutive rules} in the sense of \cite{Searle}. Section 3.2.2 extends it to an account of quantum probabilities and the Born Rule.

\subsection{Quantum Bayesianism}
The central idea of quantum Bayesianism is that quantum probabilities, as encoded in quantum states via the Born Rule, are \textit{subjective} probabilities in the subjective Bayesian sense. As such, they are construed as reflecting the state assigning agents' subjective degrees of belief as to what the results of their future ``interventions into nature''\footnote{See \cite{Fuchs} p.\ 7.} might be. Degrees of belief may legitimately differ from agent to agent without any of them behaving irrationally or making any kind of mistake. So, from the quantum Bayesian point of view different agents may indeed legitimately assign different quantum states to the same system, no notion of a quantum state as a state some quantum system ``is in'' is employed, and the position qualifies as a potentially consistent epistemic account of quantum states. According to Fuchs, ``quantum states do not exist''\footnote{Thus the title of section II of \cite{Fuchs2010}.} in that they are not part of the furniture of the world. They do not correspond to any physical facts or properties of physical system but, instead, represent the assigning agents' degrees of belief as to what the consequences of their actions on the systems the states are assigned to might be.

Quantum Bayesianism, as originally conceived, is not merely a novel philosophical perspective on quantum theory but also an ambitious programme of re-formulating the theory in terms of probabilistic and information-theoretic notions rather than abstract mathematical ones. Such an information-theoretic re-for\-mu\-lation, it is hoped, might help us understand which elements of quantum theory represent physical features of the world and which others are derived from what constitutes rational reasoning of the agents using the theory. The strategy Fuchs announces for his programme is this: ``Weed out all the terms that have to do with gambling commitments, information, knowledge, and belief, and what is left behind will play the role of Einstein's [space-time] manifold'', namely, as he explains, ``a mathematical object, the study of which one can hope will tell us something about nature itself, not merely about the observer in nature.''\footnote{See \cite{Fuchs} p.\ 6.} In what follows I shall not be concerned with the prospects and problems of this re-formulation programme for quantum theory, even though it certainly merits the attention it gets, but rather focus on the assumptions about the status of quantum observables, states and probabilities on which it is based.

One of the greatest achievements of quantum Bayesianism is its response to the challenge of not being able to make sense of talk about ``unknown quantum states'', even though such talk has a prominent role, for instance in quantum state tomography. From the quantum Bayesian point of view, the notion of an ``unknown quantum state'' makes no sense: any quantum state is always \textit{someone's} quantum state, assigned to a physical system by an agent and representing her subjective degrees of belief. The result on which quantum Bayesians ground their response is the so-called \textit{quantum de Finetti theorem}, which makes it understandable why different agents who register the same measured data generally come to agreement in their assignment of quantum states in the long run, even if the states they start out with are very different. The main presupposition used is that the agents (subjectively) judge the states of the sequence of measured systems to be exchangeable, roughly meaning that for all agents both the order of measured events and the number of measurements witnessed are irrelevant. As demonstrated by Caves, Fuchs, and Schack, this suffices to guarantee that ``the updated probability $P(\rho|D_K)$ becomes highly peaked on a particular state $\rho_{D_K}$ dictated by the measurement results, regardless of the prior probability $P(\rho)$, as long as $P(\rho)$ is nonzero in a neighbourhood of $\rho_{D_K}$''\footnote{See \cite{Caves2002b} p. 4541.} This result can be used to account for the fact that the states assigned by different agents will converge after a sufficiently large number of measurements witnessed without postulating that there is any such thing as the agent-independent unknown \textit{true} quantum state the states assigned by the different agents converge to. As this shows, taking seriously quantum theoretical practice inasmuch as it involves talk about ``unknown quantum states'' does not mean that one has to embrace an ontic account of quantum states, in which such talk is interpreted literally.

In order to be consistent as fully-fledged subjective Bayesian interpretation of quantum probabilities, quantum Bayesianism goes very far in interpreting elements of the quantum mechanical formalism as subjective. The view is particularly radical in its rejection of the innocently looking assumption that there can be an objective matter as to which observable some numerical value obtained in an experiment is a value of. To put it differently, quantum Bayesianism rejects the idea that the question of which observable is measured in which experimental setup ever has a determinate answer. In Fuchs' own wording the main motivation for this astonishing claim is the following:
\begin{quote}
Take, as an example, a device that supposedly performs a standard von Neumann measurement $\{\Pi_d\}$, the measurement of which is accompanied by the standard collapse postulate. Then when a click $d$ is found, the posterior quantum state will be $\rho_d=\Pi_d$ regardless of the initial state $\rho$. If this state-change rule is an objective feature of the device or its interaction with the system---i.\,e., it has nothing to do with the observer's subjective judgement---then the final state must be an objective feature of the quantum system. (\cite{Fuchs} p.\,39)
\end{quote}
According to this line of thought, if we suppose that the question of which observable some value obtained in an experiment is a value of has a determinate answer, applying the projection postulate to the state assigned to the system prior to measurement results in a uniquely determined post-measurement state, depending only on the observable measured and the measured result, not on the state assigned previously. But this means that assigning any other state than the one obtained from application of the projection postulate would be wrong. At least with respect to those cases where the post-measurement state has no dependence at all on the pre-measurement state (as in projective measurement of a non-degenerate observable), it seems difficult to avoid the conclusion that this must be the true quantum state of the system---the one it is in---, yet this conclusion is incompatible with a non-ontic reading of quantum states, according to which quantum states \textit{in general} are not states quantum systems are in. Even in those cases where the post-measurement state has a residual dependence on the pre-measurement state, application of the projection postulate imposes strong constraints on the post-measurement state if one assumes that which observable has been measured and which value has been obtained are objective matters. Since such constraints severely restrict admissible state assignments, quantum Bayesians, for the sake of consistency of their subjective Bayesian take on quantum probabilities, feel forced to deny that there can be an objective answer to the question of which observable some measured value is a value of.\footnote{An alternative move with the same consequences would be to accept that the question of which observable some measured value is a value of has a determinate answer while denying that which \textit{mathematical object} (which linear operator, say) represents that observable is an objective matter---even after a Hilbert space representation of the canonical commutation relations has already been chosen and other observables have already been associated with linear operators. Since this option is for all dialectical purposes equivalent to the one discussed in the text I shall not consider it any more in what follows as a possible option for the quantum Bayesian.}

This conclusion, however, is extremely difficult to swallow. If there could be no fact of the matter as to which observable some measured value is a value of according to \textit{all} versions of the epistemic conception of quantum states, one could well regard this result as a reductio of the epistemic conception of states itself. The main problems with this conclusion are, first, that in quantum theoretical practice there is virtually always agreement on which observable some value measured in some setup is a value of, and it seems difficult to imagine how quantum theory could be as empirically successful as it is if this were not the case. Second, if there were no fact of the matter as to which observable some measured value is a value of, there could never be any knowledge of the values of observables, at least not knowledge obtained in experiment. It seems clear, however, that physicists often do have knowledge of the values of at least some observables, and this makes the idea that there is no determinate answer to the question of which observable is measured in which setup even more problematic.\footnote{See \cite{Friederich}, Section 3, for a more detailed version of this argument against the quantum Bayesian claim that there can be no fact of the matter as to which observable some value obtained in experiment is a value of.} The account presented in the following subsection of this paper can be seen as an attempt to improve on quantum Bayesianism in precisely these respects.

\subsection{The Rule Perspective}
\subsubsection{Quantum state assignment}
As I have argued, quantum Bayesianism's denial of the assumption that there can be any fact of the matter as to which observable some numerical value obtained in an experiment is a value of leads to serious problems. It seems therefore worth investigating whether one really needs to deny this assumption for consistency in epistemic accounts of quantum states. In what follows I shall give a brief argument that this is not the case. In particular, as I shall argue, abandoning the notion of a quantum state that a quantum system is in does \textit{not} necessarily entail abandoning the notion of a quantum state assignment being performed correctly.

Assuming that there is an objective fact as to which observable some measured value is a value of, it seems plausible that agents having registered such a value must update the states they assign in a fixed and determinate way by applying the projection postulate, taking into account the measured result. However, even with respect to cases where this narrows down possible post-measurement states to assign to a uniquely determined (pure) quantum state, saying this is \textit{not} the same as concluding that this state is the \textit{true} post-measurement quantum state of the system after measurement, applying to the system in an agent-independent way. The simple reason for this is that other agents need not have had any chance to register the measured event due to how they are physically situated (outside the future light cone of the measurement process, say). In this case, if we take seriously the idea that quantum states should reflect the epistemic situations of the agents assigning them, assigning the same state as those who have registered the measured result would not only not be mandatory for those who haven't registered it, it would even be wrong, for it would not conform to what they know of the values of observables of the system.

Furthermore, even if one interprets the state assigned after measurement as somehow objectively privileged or distinguished, this does not mean that other (let alone \textit{all}) quantum systems must have equally privileged quantum states (to be identified with their ``true'' ones), independently of whether there happen to be any agents having knowledge of the values of their observables. The assumption that even when there are no such agents there exists some state which \textit{would have to be assigned} by anyone intending to assign correctly need not be made and does not go well with the epistemic conception of states. To conclude, admitting that there \textit{can be} (and in general \textit{is}) an objective matter as to which observable some measured value is a value of does not mean that one has to abandon the epistemic conception of quantum states.\footnote{See Section 4 of \cite{Friederich} for a defence of this line of thought against the anticipated charge that it accords too much weight to the state assignments of agents who are simply not well-informed about the measured system.}

If one wants to combine an epistemic account of quantum states with the idea that the measured observable is an objective feature of the measuring device, sense has to be made of the idea of a state assignment being performed correctly without thereby acknowledging the notion of a quantum state a quantum system ``is in''. The account proposed in \cite{Friederich} does so by appealing to the rules according to which the assignment of quantum states is performed in the application of quantum theory, arguing that to assign in accordance with them is what it means to assign correctly. An example of a rule governing state assignment is unitary time-evolution as prescribed by the Schr\"odinger equation, which in this approach is regarded not as a fact about quantum states---that their time-evolution follows the Schr\"odinger equation---but as the rule an agent must apply to the state she assigns for all times $t$ with respect to which no incoming data concerning the values of observables are registered. Other examples of such rules include the von Neumann projection postulate (and its generalisations such as L\"uders' Rule and further generalisations in terms of POVMs) and the principle of entropy maximisation, which determines the state that must be assigned by an agent, depending on what she knows of the values of its observables, if she has not assigned any state before. From this perspective, the rules of state assignment are \textit{constitutive} rules\footnote{The terminology is due to Searle. See \cite{Searle}, pp.\ 33 f., for Searle's original, much more detailed account of the notion of a constitutive rule, contrasting it with that of a \textit{regulative} rule, which both play an eminent role in Searle's philosophy of language.}, that is, they are not strategies for obtaining (decent approximations to) the true quantum states (for there are no such states). Rather, they \textit{define} the very notion of a quantum state assignment being performed correctly. Mastering these rules is a necessary requirement for being a competent user of quantum theory itself and applying them correctly is part of correctly applying quantum theory itself. Consequently, any \textit{justification} for these rules---sometimes asked for with respect to the rules of state change in criticisms of epistemic accounts of quantum states\footnote{See, for instance, \cite{Duwell} for objections to Pitowsky's ``objective Bayesian'' approach \cite{Pitowsky} of justifying the rules of state change by appeal to rational agents' betting behaviour in the context of quantum gambles.}---comes in the form of empirical support and confirmation of the theory as a whole. Thus, the rules are constitutive of what it means to correctly assign a quantum state to a quantum system, and they partially fix the empirical significance of (the elements of) the formalism of the theory a a whole.\footnote{See \cite{Friederich}, Section 6, for more details on this point.} Due to the eminent role this account evidently attributes to the rules governing state assignment, I propose to refer to it, as already announced, as the ``Rule Perspective''. The next section discusses what quantum probabilities should be claimed to be probabilities \textit{of} in the Rule Perspective.

\subsubsection{Probabilities of what?}
An account of quantum probabilities that goes very well with the account of the rules of state assignment just outlined has recently been proposed by Richard Healey in the context of his pragmatist interpretation of quantum theory \cite{Healey}. A central notion in that interpretation is that of a \textit{non-quantum magnitude claim} (``NQMC'', for short), which refers to a statement of the form ``The value of observable $A$ of system $s$ lies in the set of values $\Delta$''. Healey calls such statements ``non-quantum'', arguing that ``NQMCs were frequently and correctly made before the development of quantum theory and continue to be made after its widespread acceptance, which is why I call them non-quantum.''\footnote{See \cite{Healey} p.\ 22.} NQMCs more or less correspond to what for adherents of the Copenhagen interpretation were descriptions in terms of ``classical concepts''\footnote{See, for instance, \cite{Heisenberg} p.\ 30.}, but Healey objects against this usage of ``classical'' that it invites the misleading impression that a NQMC ``carries with it the full content of classical physics.''\footnote{See \cite{Healey} p.\ 8.} An endorsement of a NQMC is not to be construed as entailing any commitment to the view that the dynamics of the system at issue are described by classical laws of motion. In Healey's view, taken over in the Rule Perspective, NQMCs are crucial in the application of quantum theory in that they (not quantum states) have the linguistic function of describing the phenomena and regularities quantum theory is used to predict and explain.

NQMCs are naturally regarded as the bearers of quantum probabilities, derived by the Born Rule
\begin{equation}\label{Born} 
\rm{prob}_\rho(A\in\Delta)=\rm{Tr}(\rho\Pi^A_\Delta), 
\end{equation} 
where $\rho$ denotes the density operator assigned to the system and $\Pi^A_\Delta$ the projection on the span of eigenvectors of $A$ with eigenvalues lying in $\Delta$. To a first approximation, we can read this equality as attributing a probability to a statement of the form ``The value of $A$ lies in $\Delta$'', that is, to a NQMC.

The Rule Perspective can adopt this straightforward reading of probability ascriptions to NQMCs, but it has to acknowledge the no-go theorems due to Gleason, Bell, Kochen, Specker and others, which impose severe constraints on ascribing determinate values to the observables of a quantum system. A standard response to this difficulty is to interpret the Born Rule as attributing probabilities to NQMCs only inasmuch as they report on measurement outcomes. According to this view, the quantity $\rm{Tr}(\rho\Pi^A_\Delta)$ is interpreted as the probability of obtaining a value of $A$ lying in $\Delta$, in case $A$ were to be measured. This (instrumentalist) take on the Born Rule has the unappealing feature that it construes the empirical relevance of quantum theory as restricted to measurement contexts. In addition, it seems not to do justice to quantum theoretical practice, where claims about the values of observables are often considered (and Born Rule probabilities computed for them) even where these values are not determined experimentally.\footnote{See \cite{Healey} pp.\ 9-16 for a discussion of recent experiments on environment-induced decoherence involving fullerene molecules that greatly elaborates and emphasises this point. See \cite{Schlosshauer} for a helpful introduction to decoherence and a clarification of its relevance for the presently most discussed interpretations of quantum theory.} This observation makes the challenge of clarifying the appropriate range of applicability of the Born Rule even more pressing.

Healey's pragmatist interpretation of quantum theory responds to it in a way that arguably does justice to quantum theoretical practice. Appealing to environment-induced decoherence, it holds that an agent is entitled to apply the Born Rule to just those NQMCs which refer to observables for which taking into account the system's interaction with its environment renders the density operator assigned to the system (at least approximately) diagonal. In the words of Healey:
\begin{quote}
Born-rule probabilities are well-defined only over claims licensed by quantum theory. According to the quantum theory, interaction of a system with its environment typically induces decoherence in such a way as (approximately) to select a preferred basis of states in the system's Hilbert space. Quantum theory will fully license claims about the real value only of a dynamical variable represented by an operator that is diagonal in a preferred basis: it will grant a slightly less complete license to claims about approximately diagonal observables. All these dynamical variables can consistently be assigned simultaneous real values distributed in accordance with the Born probabilities. So there is no need to formulate the Born rule so that its probabilities concern only measurement outcomes. (\cite{Healey} p.\ 15)
\end{quote}
Healey spells out what he means by saying that a NQMC is ``licensed'' by quantum theory in terms of an inferentialist account of linguistic meaning that regards the content of NQMC as determined by what ``material inferences''\footnote{See \cite{Healey} p.\ 13. Section 3 of that paper spells out Healey's account of what it means for a NQMC to be licensed by quantum theory in far more detail.} an agent applying quantum theory is entitled to draw from it. For the purposes of the present investigation I propose a slightly simplified picture according to which we may think of a NQMC of the form ``The value of $A$ lies in $\Delta$'' as ``licensed'' by quantum theory just in case an agent who applies quantum theory to the system in question is entitled to assume as the basis of her further reasoning that the value of $A$ either determinately lies within $\Delta$ or outside $\Delta$.\footnote{To what degree this simplified way of spelling out what it means for a NQMC to be ``licensed'' by quantum theory still captures the essence of Healey's more subtle inferentialist account seems difficult to assess. It is not impossible that the Rule Perspective, as presented here, may require refinement in this respect, but the present formulation seems sufficient to defend the Rule Perspective against the objections discussed in the following sections.} In accordance with Healey's claim quoted above the Rule Perspective can now say that quantum theory ``licenses'' those NQMCs which ascribe values to observables $A$ for which the (reduced) density operator $\rho$ an agent assigns to the system is at least approximately diagonalised in a preferred way by the spectral decomposition of $A$.\footnote{It seems natural to assume that for degenerate observables an (approximately) block-diagonal form of the density matrix assigned should suffice for licensing application of the Born Rule. More detailed empirical investigations as to what is conceived of as a legitimate application of the Born Rule in quantum theoretical practice would be useful to say more on this matter.} The Rule Perspective concurs with Healey's pragmatist interpretation in that it construes Born Rule probabilities as ``well-defined only over claims licensed by quantum theory'' in precisely that sense.

Environment-induced decoherence becomes relevant here in that taking into account the system's coupling to its environment and performing the trace over the environmental degrees of freedom typically makes $\rho$ (at least approximately) diagonal in an environment-selected basis. In conditions which function as measurement setups the role of the environment is typically played by the system which is used as the measurement apparatus, and the measurement fulfils its purposes if an eigenbasis of the observable meant to be ``measured'' coincides with the Hilbert space basis selected by decoherence. This accounts for the fact that quantum theory, when employed with competence, licenses application of the Born Rule in measurement contexts and thus yields probabilities for the possible values of the observable(s) the apparatus is meant to measure.

As already remarked, Healey argues for his take on the Born Rule on the basis of an inferentialist theory of conceptual content. Inasmuch as it is simply an accurate statement of what counts as correct employment of the Born Rule in practice it is independent of any particular philosophical view of linguistic meaning and content. Since the main motivation for the Rule Perspective is to dissolve the paradoxes by paying attention to quantum theoretical practice (where the paradoxes do not seem to arise as practical difficulties), it is natural to combine it with Healey's account of the Born Rule as applying to those NQMCs for which environment-induced decoherence makes the density matrix assigned (at least approximately) diagonal. These NQMCs may differ from agent to agent, so not only the values of quantum probabilities depend on the assigning agent, but also what their bearers are.

\section{Challenging the interpretation of probabilities}
Quantum Bayesianism construes quantum probabilities as subjective in the subjective Bayesian sense of individual agents' degrees of belief. Using the subjective/objective distinction to characterise the status of quantum probabilities in the Rule Perspectiveis is less straightforward. On the one hand, they are ``subjective'' in that different agents (``subjects'') should ascribe different probabilities to one and the same NQMC if the agents' epistemic conditions differ. On the other hand, quantum probabilities are ``objective'' in that for sufficiently specified epistemic conditions the probability to be ascribed to a NQMC is completely fixed by the rules governing state assignment together with the Born Rule.

One widely acknowledged feature of objective probabilities is that they impose constraints on rational degrees of belief. David Lewis, for instance, referring to objective probabilities as ``chances'', contends that one should not ``call any alleged feature of reality `chance' unless [one has] already shown that [one has] something, knowledge of which could constrain rational credence.''\footnote{See \cite{Lewis94}, p.\ 484} Lewis' famous \textit{Principal Principle} requires that an agent's rational credences should be equal to the chances, provided the agent has epistemic access to them.\footnote{Lewis includes a proviso here, which requires that the agent does not have any evidence that qualifies as ``inadmissible'', i.e. such that if one has such evidence, it is no longer rational to align one's credences with the chances. I hope to say more about what should count as (in)admissible evidence in the quantum theoretical context in a future publication.} It is natural to conceive of the Rule Perspective as saying that agents using quantum theory may employ the Principal Principle as a guide when forming their credences about NQMCs in the light of the Born Rule probabilities derived from the quantum states they assign. Hence, the Rule Perspective regards quantum probabilities as \textit{objective} inasmuch as they impose \textit{objectively valid} constraints on rational credences about NQMCs; it regards them as \textit{subjective} inasmuch as what the correct probabilities are depends on the evidence the agent has access to.

There are certain limitations to the applicability of the Principal Principle in the context of the Rule Perspective which should be noted. They concern situations where an agent has to take into account the possibility that the system at issue might be affected in ways that can undermine her knowledge about which NQMCs are true. In the EPR-Bohm setup, for instance, where Alice performs her measurement of $S_z$ and obtains the result $+1/2$, the rules governing state assignment commit her to assign the state $|\downarrow_z\rangle$ to Bob's system for times immediately after that measurement. This state ascribes probability $0$ to the value of $S_z$ of Bob's system being $+1/2$. However, if Bob decides to measure first spin in another direction, say $S_x$, and to measure $S_z$ only subsequently, he may well obtain the result $+1/2$ in this latter measurement of $S_z$. Alice should no doubt account for this possibility in her credences. So, with respect to these cases her rational credences cannot simply be identified with the probabilities derived from the state $|\downarrow_z\rangle$, even though the Principal Principle might seem to suggest this.

\subsection{The means/ends objection}

Interpretations of quantum probabilities as subjective have been criticised on a number of grounds. Since both accounts discussed here conceive of quantum probabilities as subjective in at least the (restricted) sense of being agent-relative, it makes sense to consider two especially pointed (and closely related) criticisms of accounts of quantum probabilities as subjective. Both were originally formulated by Chris Timpson as objections against quantum Bayesianism. My conclusion will be that despite the important points of agreement between the two positions, the objections apply only to quantum Bayesianism.

The objection I consider first is referred to by Timpson as the \textit{means/ends objection}. Its underlying idea is that any account which denies quantum probabilities the status of objective features of the world inevitably makes it mysterious how the theory helps us with as little as ``the pragmatic business of coping with the world''---let alone with the more ambitious goal of ``finding out how the world is.''\footnote{See \cite{Timpson} p.\ 606.} According to Timpson, any interpretation of quantum probabilities that does not conceive of them as objective single case probabilities makes it unclear why updating our assignments of probabilities in the light of new data should be useful and enhance our predictive success: ``[I]f gathering data does not help us track the extent to which circumstances favour some event over another one (this is the denial of objective single case probability), then why does gathering data and updating our subjective probabilities help us do better in coping with the world?''\footnote{Ibid. p.\ 606.} What makes Timpson's worries most pressing is that for the quantum Bayesians the question of \textit{in which way} an agent should update her probability assignments in the light of new data does not have an objective answer. The reason for this is that quantum Bayesianism does not acknowledge any fact of the matter as to which observable is being measured in which setup and what update of one's state assignment one should make. On this view, the predictive and pragmatic success of quantum theory---why it helps us ``coping with the world''---is indeed mysterious: if there is no objective answer as to \textit{how} some assignment of probabilities should be updated in the light of new evidence, it becomes a miracle that updating probabilities is of use at all. The main force of the means/ends objection against quantum Bayesianism, as we see, derives from the quantum Bayesian's claim that we can never know what observable some measured value is a value of.

Timpson outlines a possible quantum Bayesian reply to this challenge, which, as he says, is ``of broadly Darwinian stripe''. According to this reply, ``[w]e just do look at data and we just do update our probabilities in light of it; and it is just a brute fact that those who do so do better in the world; and those who do not, do not.''\footnote{Ibid. p.\ 606.} However, as he argues, this response is ultimately unsatisfying in that it does not address the original worry, namely, the nagging question ``why do those who observe and update do better[.]''\footnote{Ibid. p.\ 606.} Given that, for quantum Bayesianism, no way of updating in the light of incoming data counts as correct (in contrast to all others ways one might think of), the challenge is particularly serious.

However, as a moment's reflection makes clear, this problem is not generic in epistemic accounts of quantum states. The Rule Perspective, for instance, avoids it. To see this, recall that according to the Rule Perspective quantum theory helps us determine and predict which \textit{non-quantum} claims (NQMCs) are true. So, on this view the theory is not only a tool that helps us accomplish the ``business of coping with the world'' but also one that helps us with the more ambitious goal of ``finding out how the world is'', to use Timpson's words. In contrast with quantum Bayesianism, the Rule Perspective concedes that we often do have knowledge of the values of observables, and it regards the practice of making probability ascriptions and updating them in the light of new evidence as directed at the aim of improving that knowledge in various (direct and indirect) ways. The Rule Perspective does not deny that an intimate connection exists between objective features of the world, as described in terms of NQMCs, and quantum state assignments, performed on grounds of what NQCMs one knows to be true. Therefore, it is not miraculous that updating our state assinments,---and with them our probability ascriptions---can help us predict which NQMCs are true. There remains no ``explanatory gap''\footnote{Ibid. p.\ 606.}, as Timpson objects to quantum Bayesianism, between the methods of enquiry---assigning quantum states and deriving probabilities from them---and the goals it seeks to achieve---broadly (and somewhat crudely), to determine which NQMCs are true. Furthermore, since knowledge about ``how the world is'' is plausibly helpful for our competence in ``coping with the world'', it is small wonder that quantum theory helps us with the latter if it helps us with the first. 

\subsection{The quantum Bayesian Moore's paradox} 
The second of Timpson's criticisms against the interpretation of quantum probabilities as subjective focuses on assignments of probability $1$. According to Timpson, if one conceives of quantum probabilities as subjective degrees of belief, this commits one to the systematic endorsement of pragmatically problematic sentences of the ``quantum Bayesian Moore's paradox'' type. Sentences of this type are cousins of the better known ``Moore's paradox'' sentences, invented by G. E. Moore, which are characterised by having the form 
\begin{quote} 
$p$, but I don't believe that $p$. 
\end{quote} 
There is a long-standing philosophical debate on the status and proper interpretation of these sentences, in particular as to whether they involve a pragmatic or even a semantic contradiction, but there seems to be agreement on their paradoxical nature in that, as expressed by Timpson, they ``violate the rules for the speech act of sincere assertion.''\footnote{See \cite{Timpson} p.\ 602.} Timpson argues that by interpreting quantum probabilities as subjective degrees of belief and denying that there is such a thing as \textit{the} quantum state a quantum system is in, quantum Bayesianism is committed to the systematic endorsement of sentences having a similar structure and a similar paradoxical flavour. The problem he diagnoses occurs in connection with the assignment of quantum states ascribing probability $1$ to a possible value of at least one observable. Typical examples arise from the assignment of \textit{pure} quantum states. These ascribe only probabilities $0$ or $1$ to the possible values of observables they are eigenstates of. The problem can be seen by considering an agent who consciously accepts the quantum Bayesian take on quantum probabilities as subjective degrees of belief and assigns a pure quantum state to a system, for instance the state $|\uparrow_z\rangle$ for the spin degree of freedom of a spin-$1/2$ system. Such an agent, according to Timpson, ``must be happy to assert sentences like: `I assign a pure state (e.\ g. $|\uparrow_z\rangle$) to this system, but there is no fact about what the state of this system is.'\,''\footnote{Ibid. p.\ 604.} In other words, any quantum Bayesian agent is committed to the systematic endorsement of sentences of the form:
\begin{quote}
``\textbf{QBMP:} `I am certain that $p$ (that the outcome will be spin-up in the $z$-direction) but it is not certain that $p$.'\,'' (\cite{Timpson} p.\ 604. The acronym ``QBMP'' stands for ``quantum Bayesian Moore's paradox''.)\footnote{The analogy to Moore's original sentence ``$p$, but I don't believe that $p$'' can be made more transparent if the part of the QBMP which reports on the agent's epistemic conditions is put second, just as in Moore's original sentence. (I would like to thank Jeremy Butterfield for pointing this out to me.) A formulation that fulfils this requirement is: ``It is uncertain whether $p$, but I am not uncertain whether $p$ (that is, I am absolutely certain that $p$).''}
\end{quote}
For a quantum Bayesian, an ascription of probability $1$ to the value of an observable signals complete certainty as to what the outcome of measurement of that observable will be, but her subjective Bayesian take on probabilities (including probability $1$) implies that she cannot claim that there is any \textit{objective} certainty as to what the measurement outcome will be since she cannot countenance any ``fact determining what the real state is.''\footnote{Ibid. p.\ 605.} Sentences of the form of the QBMP seem pragmatically problematic since expressing absolute certainty seems irrational if one does not believe it to be grounded in facts. Timpson notes that similar-structured paradoxical features of ascriptions of probability $1$ (and, one might add, of ascriptions of probability $0$) are generic in accounts that are based on the subjective Bayesian take on probability and arise not only in the context of quantum Bayesianism. However, whereas those who hold subjective Bayesian views with respect to other contexts are in principle free to refrain from making extremal probability assignments, quantum Bayesians must make them unless they are prepared to abandon the assignment of pure states. As Timpson notes, ``the occurrence of these paradoxical sentences isn't just an occasional oddity which can be ignored'', but one which ``arises whenever one finds a quantum Bayesian who is happy to assign pure states and is also explicit about what their understanding of the quantum state is.''\footnote{Ibid. p.\ 605.}

There is a further difficulty for quantum Bayesianism here, which Timpson does not mention and which seems no less severe. It arises again from the quantum Bayesians' denial that we can ever have any knowledge of the values of observables or, what comes to the same, that we can ever have any knowledge as to which NQMCs are true of which system. According to quantum Bayesianism, we can never know that some NQMC ``$p$'' is true, not even in cases where we assign pure states, which, on the quantum Bayesian reading, invariably signal our being certain that ``$p$'' is true. It is difficult to see how one might coherently claim that agents cannot \textit{know} that $p$ but be \textit{certain} that $p$.

Timpson considers a possible quantum Bayesian reaction to this problem, namely, to adopt a perspective on Born Rule probability ascriptions that is similar to that of ethical non-cognitivists on moral discourse. In analogy to how non-cognitivists may aspire to explain why endorsing moral claims can be rational and legitimate despite the claimed non-existence of moral facts, quantum Bayesians might ``elaborate on how there can be a role for personal certainty within our intellectual economy which is insulated against the absence of any impersonal certainty.''\footnote{Ibid. p.\ 606.} The quantum Bayesian may try to defend her position along similar lines against the challenge sketched in the previous paragraph by suggesting that one can indeed be coherently certain that $p$ while at the same time---on a meta- or higher level---regarding it as impossible to know that $p$. However, this move appears still rather far-fetched and contrived.

Unlike quantum Bayesianism, the Rule Perspective does not conceive of quantum probabilities as \textit{expressing} their assigning agents' degrees of belief. However, it acknowledges that the Born Rule probabilities one ascribes often do have ramifications for one's rational credences. Agents who accept the Rule Perspective and apply the Principal Principle in cases where the states they assign deliver Born Rule probability $1$ for a NQMC ``$p$'' will indeed be certain that $p$. It may therefore seem that the proponent of the Rule Perspective has the same problems as the quantum Bayesian with respect to QBMP-type sentences in that both are committed to the systematic endorsement of sentences of the form ``I am certain that $p$, but it is not certain that $p$''.\footnote{I would like to thank an anonymous referee of \textit{Studies in History and Philosophy of Modern Physics} for suggesting the problem in this form to me.}

However, the Rule Perspective differs from quantum Bayesianism in that it acknowledges that whether a NQMC ``$p$'' is true is something we can (under favourable circumstances) \textit{know}. There is therefore no reason for an adherent of the Rule Perspective to deny that her assignment of probability $1$ to some NQMC is related to a fact she has evidence of, namely, the fact that things are indeed as described by ``$p$''. Inasmuch as impersonal certainty is construed as entailed by truth (that is, if ``it is \textit{true} that $p$'' is regarded as entailing ``it is certain that $p$''), this removes the problem for the Rule Perspective. On \textit{this} reading of ``it is certain that $p$'', whenever an agent is entitled to be certain that $p$ (by applying the Principal Principle to the Born Rule probabilities derived from a state she assigns on the basis of the rules governing state assignment) she may well endorse the claim ``I am certain that $p$ \textit{and it is certain} that $p$''. So, the QBMP does not arise.

In response to this line of thought critics may propose a different reading of ``it is certan that $p$'', according to which impersonal certainty is not entailed by truth and according to which the adherent of the Rule Perspective may seem to be committed to the endorsement of QBMP-type sentences after all. One may argue that being \textit{rationally} certain that $p$ requires being able to exclude that there are bits of evidence which, if one had access to them, would make one stop being certain that $p$. This requirement may appear to be violated in the Rule Perspective, where an ascription of probability $1$ must potentially be revised (and the certainty that may go with it be abandoned) whenever new relevant evidence about which NQMCs are true is obtained. It may therefore seem that agents who accept the Rule Perspective and perform an ascription of Born Rule probability $1$ are committed to a statement of the following form (where ``MQBMP'' stands for ``modified quantum Bayesian Moore's paradox'')
\begin{quote}
I am certain that $p$, but there may be additional evidence which, if I had it, would make me stop being certain that $p$. (MQBMP)
\end{quote}
Irrespective of how problematic such a statement really is, it would probably not be good news for the Rule Perspective if its adherents were committed to the systematic endorsement of it.

Fortunately, this is not the case. To see why, it is useful, first, to recall that the Rule Perspective regards entropy maximisation as a constitutive rule of state assignment (to be applied whenever a quantum states is assigned to a system for the first time) and, second, to observe that entropy maximisation helps avoid situations where an agent must revise an ascription of probability $1$ in the light of additional evidence. As explained by Jaynes, what distinguishes an entropy-maximising probability function from its alternatives is that ``no possibility is ignored; it assigns positive [probability] weight to any situation that is not absolutely excluded by the given information.''\footnote{See \cite{Jaynes} p.\ 623.} Similarly, what distinguishes entropy-maximising quantum states from \textit{non}-entropy-maximising ones is that entropy-maximising ones ascribe probability \textit{strictly less than one} to any NQMC ``$p$'' which is not entailed by the information about true NQMCs on grounds of which the state assignment is made.

As a consequence, whenever an agent obtains information about a quantum system that requires revision of an ascription of Born Rule probability $1$, she will interpret it as pertaining to the system in a \textit{novel physical situation}, where the information about true NQMCs which she used to make her original probability ascription is no longer valid. For example, if an assignment of $|\uparrow_z\rangle$ is made on the basis of knowledge that the NQMC ``The value of $S_z$ is $+1/2$'' is true, then, if  $S_x$ is measured, the ascription of probability $1$ to that NQMC must be revised. The revised probability ascription, however, must be read as pertaining to the system \textit{in a different physical situation} than before. Consequently, whenever an adherent of the Rule Perspective is certain, on grounds of a correctly performed state assignment, that some NQMC ``$p$'' holds true of the system at $t$, she can also be certain that any additional evidence which, if she had it, would lead her not to be certain that ``$p$'' concerns the system at a different time $t'\neq t$. Thus, an adherent of the Rule Perspective would not endorse any statement of the form (MQBMP) with both occurrences of ``$p$'' referring to the same system at the same time. To conclude, it seems that the quantum Bayesian Moore's paradox does not arise as a problem for the Rule Perspective on any of the paradox's possible readings.

\section{Challenging anthropocentric notions} 
\subsection{Bell's criticism}
Quantum Bayesianism and the Rule Perspective are formulated in terms of notions such as ``agent'', ``epistemic situation'' and ``state assignment'', which are neither themselves fundamental physical notions nor in any evident way reducible to such notions. In the present section I discuss an objection against the employment of such ``\textit{anthropocentric} notions'' (as I will call them), which claims that they are far too vague and too imprecise to be used in foundational accounts.

First, let us have a closer look at the role of anthropocentric notions in epistemic accounts of quantum states. The notion of an \textit{agent}, to begin with, is employed in the statement of the epistemic conception of quantum states itself, claiming that quantum states reflect the assigning agents' epistemic conditions with respect to the system states are assigned to. Clearly, this idea cannot be expressed without using some notion of a subject who has an epistemic condition and assigns a quantum state. Furthermore, ``epistemic condition'' and ``state assignment'' are themselves no less anthropocentric than ``agent'' (or ``subject''). Both appear equally irreducibly and ineliminably in the statement of the epistemic conception of states itself. In addition, some relative of the notion of measurement is employed in both quantum Bayesianism and the Rule Perspective: In the case of quantum Bayesianism the relevant notion is that of an ``experimental intervention[...] into nature''\footnote{See \cite{Fuchs} p.\ 7 and various other places.}; in the case of the Rule Perspective it is that of an event resulting in ``new knowledge'' of the values of observables, to be taken into account when updating one's state assignment in accordance with L\"uders' Rule. Different epistemic accounts of quantum states may use different anthropocentric notions, but in view of the crucial role which anthropocentric notions play in the statement of the epistemic conception of states itself and in the more detailed considerations of the two versions discussed here it seems highly plausible that they cannot dispense with them altogether. 

Interpretations of quantum theory that rely on anthropocentric notions are heavily criticised by some of the most distinguished interpreters of the theory. J. S. Bell, for instance, claims that such notions are not sufficiently sharp and fundamental to be used in foundational accounts. Among the words which, according to him, ``however legitimate and necessary in application, have no place in a \textit{formulation} with any pretension to physical precision''\footnote{See \cite{Bell} p.\ 209. The emphasis is Bell's.} one finds, for instance, ``observable, information, measurement'', which, evidently, are close relatives of the anthropocentric notions encountered in the epistemic accounts of quantum states discussed here. Bell's main complaint concerns the role of these notions in accounts which postulate the occurrence of measurement collapse whenever a quantum system is measured. Famously, Bell comments sarcastically on this idea: 
\begin{quote} 
What exactly qualifies some physical systems to play the role of `measurer'? Was the wavefunction of the world waiting to jump for thousands of millions of years until a single-celled living creature appeared? Or did it have to wait a little longer, for some better qualified system ... with a PhD? If the theory is to apply to anything but highly idealised laboratory operations, are we not obliged to admit that more or less `measurement-like' processes are going on more or less all the time, more or less everywhere? Do we not have jumping then all the time? (\cite{Bell} p.\ 34.) 
\end{quote} 
Bell's criticism of the textbook picture of measurement collapse as occurring when\-ever the system is measured seems very reasonable and intuitive. However, it does not directly apply to the versions of the epistemic conception of quantum states considered here. These deny that there is such a thing as \textit{the} wavefunction of a quantum system with respect to which the question of when it ``jumps'' could be meaningfully asked. In particular, according to these accounts there exists no such thing as a ``wavefunction of the universe'', which at one point ``jumps'' for the first time in its history. Primitive anthropocentric notions are indeed used, but their role is not that of specifying under which conditions which physical process takes place but to characterise, on a conceptual level, the elements of the formalism of quantum theory as employed by competent physicists.

The difference between these two different kinds of appeal to anthropocentric notions is clarified by Fuchs in his quantum Bayesian characterisation of quantum theory as a ``users' manual that \textit{any} agent can pick up and use to help make wiser decisions in this world of inherent uncertainty.''\footnote{See \cite{Fuchs2010} p.\ 8.} In this picture of quantum theory, the theory is construed as an empirical extension of subjective Bayesian probability theory, where probability theory, in turn, is construed as an ``extension of logic.''\footnote{See \cite{Fuchs2010} fn.\ 14 on p.\ 8.} Fuchs argues convincingly that on this conception quantum Bayesianism cannot be plausibly asked to deliver a reductive analysis of notions such as ``agent'' any more than philosophers of logic can be asked to deliver a reductive account of a notion such as ``logical subject'' (that is, the notion of an agent who employs the methods of logic):
\begin{quote} 
[I]s ... quantum mechanics ... obligated to derive the notion of agent for whose aid the theory was built in the first place? The answer comes from turning the tables: Thinking of probability theory in the personalist Bayesian way, as an extension of formal logic, would one ever imagine that the notion of an agent, the user of the theory, could be derived out of its conceptual apparatus? Clearly not. How could you possibly get flesh and bones out of a calculus for making wise decisions? The logician and the logic he uses are two different substances---they live in conceptual categories worlds apart. One is in the stuff of the physical world, and one is somewhere nearer to Plato's heaven of ideal forms. Look as one might in a probability textbook for the ingredients to reconstruct the reader himself, one will never find them. So too, the Quantum Bayesian says of quantum theory. (\cite{Fuchs2010} p.\ 8 f.) 
\end{quote} 
Fuchs' point is that if quantum theory is conceived of as normative, namely, as a ``manual'' to ``make wiser decisions'', one can hardly expect that the notion of an agent applying the methods of that manual might be spelled out in terms of the notions which the manual uses itself. Thus, Bell's criticism of the employment of anthropocentric notions in interpretations of quantum theory does not seem to apply to the epistemic accounts of quantum states considered here due to \textit{how} they rely on such notions. Fuchs' comparison between quantum theory and logic might be criticised on grounds that quantum theory, unlike logic, is a physical theory and that agents using quantum theory and the world they live in are themselves (aggregates of) objects studied in physics, whereas agents availing themselves of the methods of logic and the world they live in are not (aggregates of) objects studied in logic (whatever exactly one takes these to be). The crucial point of Fuchs' argument, however, is that in quantum Bayesianism anthropocentric notions are employed at the \textit{meta-level} of characterising the status of quantum theoretical concepts---claiming that quantum states are used non-descriptively, for instance---not at the \textit{object-level} of describing physical processes themselves---such as physical collapse and under which conditions it occurs. The textbook accounts criticised by Bell, in contrast, make object-level use of anthropocentric notions in that they conceive of measurement collapse as a physical process that takes place whenever the system is measured.

To sum up, Bell's verdict against anthropocentric notions in foundational accounts seems reasonable only with respect to accounts which invoke these notions at the object-level of physical processes. There is no reason to extend it to accounts which use anthropocentric notions to clarify the status and linguistic role of the elements of the quantum theoretical formalism. Inasmuch as this is what quantum Bayesianism and the Rule Perspective do, they are not concerned by Bell's criticism. The following subsection investigates whether epistemic accounts of quantum states can confine their appeal to anthropocentric notions to the meta-level of conceptual clarification even when when they try to answer the challenge of why those observables which we ``measure'' have always determinate values.

\subsection{Anthropocentric notions and value determinateness}
In debates about the foundations of quantum theory it is often claimed that appeal to ``measurement'' should not be understood as entailing the existence of a value of the quantity taken to be ``measured'' prior to the measurement act. The idea that a quantum measurement is an act of \textit{creation} rather than an act of establishing what is there has a long tradition. Pascual Jordan, for instance,endorses it when he writes that it is ``we ourselves [who] bring about the matters of fact''\footnote{See \cite{Jordan} p.\ 228, my translation.} which we usually think of as being determined in experiments. Similar ideas can be found in the writings of adherents of the epistemic conception of quantum states, for instance in those of the quantum Bayesians Fuchs and Schack, who claim that the ``measured values'' of observables are not merely ``\,`read off'\,'' in measurement but rather ``enact[ed] or creat[ed] ... by the [measurement] process itself.''\footnote{See \cite{FuchsSchack2} p.\ 3.}

Quantum Bayesianism's main motivation for denying the existence of determinate values prior to measurement is the conflict between assuming such values and the famous no-go theorems on determinate value assignments originating from Gleason, Bell, Kochen and Specker. As explained in Section 3.1, quantum Bayesianism denies that we can ever know which observable some numerical value obtained in an experiment is a value of, but it doesn't go as far as claiming that observables do not have any values at all. As a result, quantum Bayesianism is drawn to the conclusion that determinate values are created in the act of measurement. In the words of Fuchs:
\begin{quote}
QBism [i.\ e., quantum Bayesianism as defended by Fuchs] says when an agent reaches out and touches a quantum system---when he performs a quantum measurement---that process gives rise to birth in a nearly literal sense. With the action of the agent upon the system, the no-go theorems of Bell and Kochen-Specker assert that something new comes into the world that wasn't there previously: It is the ``outcome,'' the unpredictable consequence for the very agent who took the action. (\cite{Fuchs2010} p.\ 8.)
\end{quote}
And a few pages later:
\begin{quote}
That the world should violate Bell's theorem remains, even for QBism, the deepest statement ever drawn from quantum theory. It says that quantum measurements are moments of creation. (\cite{Fuchs2010} p.\ 14)\footnote{See the rest of Section V of \cite{Fuchs2010} for a detailed argument using Bell-Kochen-Specker-type reasoning for the view that measurements bring about the values of observables determined through them.}
\end{quote}
The problem with this conception of ``quantum measurement'' as a type of ``birth'' or ``creation'' is that it seems to lead straightforwardly to an application of ``measurement'' in what according to the terminology used in the previous subsection counts as an \textit{object-level} type of way. In this case, it is not the occurrence of collapse as a physical process which is taken to occur when a measurement is performed, but the ``birth'' of the value of an observable when it is measured. According to these considerations, it seems that quantum Bayesianism is unable to confine its employment of anthropocentric notions to the meta-level of conceptual clarification alone, in which case Bell's criticism applies after all.

From a general perspective, the challenge of accounting for determinate post-measurement values without relying on object-level anthropocentric notions can be construed as a re-appearance of the measurement problem in disguise: As explained in Section 2.1, the original measurement problem is dissolved in epistemic (or, more generally, non-ontic) accounts of quantum states by rejecting the notion of the quantum state a quantum system is in, but the question as to why measurement processes, as a matter of fact, do always result in determinate outcomes remains unaddressed. It is tempting to read the passages of the quantum Bayesians on ``quantum measurement'' as a type of ``birth'' or ``creation'' as a response to this challenge that invokes primitive anthropocentric notions at the object-level. In that case, however, Bell's criticism applies to it and its dissolution of the measurement problem cannot be considered successful.

In defence against this charge, quantum Bayesians may point out that to conceive of measurement as an act of ``birth'' of the value of an observable is not the same as using ``measurement'' to \textit{define} under which conditions determinate values are assumed. According to Fuchs, ``for the QBist, the real world ... ---with its objects and events---is taken for granted.''\footnote{See \cite{Fuchs2010} p.\ 7.} Quantum Bayesianism, one may take this to be saying, \textit{presupposes} the existence of ``events'' where determinate values are assumed as a primitive fact, which is not in need of any further grounding. They might add that among these events are the ``measurement events''---namely, those which we use to obtain information as to what the values of observables really are. Phrased differently (and perhaps somewhat crudely): Observables do not assume determinate values \textit{because} some process is a measurement process (an idea analogous to the one ridiculed by Bell), but we \textit{call} certain processes measurement processes because the values of observables taken to be ``measured'' are determinate at their end (in such a way that we can obtain information about them). Following this line of thought, the quantum Bayesian account of ``measurement'' as an act of ``birth'' does not necessarily lead to the employment of anthropocentric notions at the object-level.

What quantum Bayesianism does not account for, however, is how physicists can be \textit{confident} as regards under which conditions which observables have determinate values. The Rule Perspective is in better shape than quantum Bayesianism here in that it construes quantum theory as entailing that determinate values of the observables meant to be measured are precisely what competent users of quantum theory \textit{should expect}.

To see this, it suffices to conceive of a measurement setup schematically as a two-part system, which consists of a ``measured system'' $S_1$ on the one hand and a ``measurement apparatus'' $S_2$ on the other, such that information about the value of the ``measured observable'' $A$ (pertaining to the measured system) can be gained from registering the value of the ``pointer'' observable $X$ (pertaining to the apparatus and discriminating between different display configurations). Now, what it takes for a system to qualify as a candidate ``measurement setup'' is that effects from environmental decoherence, when they are taken into account, will render the density matrix assigned to the combined system $S_1\cup S_2$ approximately diagonal in an eigenbasis of $A\otimes X$. As the Rule Perspective claims (following Healey, see Section 3.2.2), quantum theory in this case ``licenses'' (Healey's wording) various NQMCs which state that the values of $A$ and $X$ lie in ranges $\Delta_A$ and $\Delta_X$ of possible values the observables $A$ and $X$ might assume. From the point of view of the Rule Perspective, to treat a NQMC of the form ``The value of $A$ lies in $\Delta_A$'' as ``licensed'' by quantum theory means to proceed by assuming that the value of $A$ lies either (determinately) within $\Delta_A$ or outside $\Delta_A$. Quantum theory thus gives the agent an entitlement to treat the values of both the ``measured'' and the ``pointer'' observable as determinate (within bounds as determined by decoherence), so operating under the assumption that these observables do \textit{not} have determinate values (whatever this would practically mean) would mean \textit{failure} to apply the theory correctly.

The Rule Perspective thus not only dissolves the measurement problem by undermining its formulation in terms of quantum states quantum systems ``are in'' but allows the further claim that outcome determinateness is to be presupposed by competent users of the theory. If an experimental setup, designed to measure the value of $A$, is such that taking into account decoherence effects does \textit{not} provide an entitlement to treat the value of $A$ as determinate, the setup simply does not qualify as a candidate ``measurement setup'' for $A$. Verifying that the setup qualifies as a ``measurement setup'' for $A$, in other words, is the same as ascertaining that one has an entitlement to assume that the value of $A$ is determinate in it. Without emplyoment of primitive anthropocentric notions at the object-level this accounts for why outcome determinateness is something that physicists have to assume in order to apply quantum theory correctly.

\section{Challenging explanation without ontic quantum states}
\subsection{The micro/macro divide}
One of the defining aspects of the epistemic conception of quantum states is that it conceives of quantum states as non-descriptive. In this section, I consider the criticism that \textit{any} non-descriptive reading of quantum states entails an unattractive ontological quantum/classical divide. This divide is supposed to separate a realm of classical macro-objects, which are susceptible to descriptions in terms of classical concepts, from a realm of quantum micro-objects, for which no descriptive account can be given at all.

Timpson makes the claim that such an ontological divide is a natural consequence of quantum Bayesianism, but does not regard this as a devastating objection against that position. To substantiate the claim that the idea of an ontological quantum/classical divide is coherent, he proposes an ontological framework which comprises, on the one hand, a ``micro-level we have dubbed unspeakable to which we are denied direct descriptive access''\footnote{See \cite{Timpson} p.\ 597.} and, on the other hand, a ``macroscopic or classical level [which] will be a level of objects which do have unproblematically stateable truths about them.''\footnote{Ibid. p.\ 598.}

Any such two-level ontological framework is confronted with obvious difficulties: is the idea that reality is a patchwork which consists of a describable and an undescribable (``unspeakable'') ontological level really intelligible? And, if it is, where exactly should the line be drawn which separates the two levels? Drawing on ideas from Nancy Cartwright's philosophy of science, Timpson suggests that the framework can indeed be made coherent if one admits at the macro-level ``metaphysically emergent properties [which] a composite can possess but which its components cannot[,] and which are not conferred on it by the properties possessed by its components and the laws (if any) which they obey.''\footnote{Ibid. p.\ 599.} As he seems to suggest, accepting such properties and their anti-reductionist ramifications is the price to be paid for a non-descriptive reading of quantum states. If the quantum Bayesian is willing to pay this price, the challenge can be met: ``[i]f called upon, ... the quantum Bayesian seems able to present an intelligible ontology to underlie their position.''\footnote{Ibid. p.\ 600.}

Others are more sceptical as regards the intelligibility of a two-level ontology comprising a level of objects that admit a descriptive account and a level of objects which don't. Marchildon, for instance, regards the empirically manifest fact that macroscopic objects ``are always in definite states''\footnote{See \cite{Marchildon} p.\ 1462.} as inexplicable on an epistemic account of quantum states. According to him, these accounts have to choose between three equally unattractive claims, namely, that either microscopic quantum objects ``do not exist'', that they ``may exist, but have no states'' or that they ``may exist and may have states, but attempts at narrowing down their existence or specifying their states are useless, confusing, or methodologically inappropriate.''\footnote{Ibid. p.\ 1462.} As he sees it, all these options are implausible, given that the ``unspeakable'' (Timpson's expression) micro-objects are the (mereological) constituents of the macro-objects for which unproblematic ``definite states'' do exist. Whether or not one regards this criticism as ultimately compelling, the challenge of combining a macro-level of describable ``classical'' objects and a micro-level of ``unspeakable'' quantum objects in a coherent metaphysical framework remains formidable. However, as I shall argue now, the example of the Rule Perspective shows that a two-level ontology of this kind need not be embraced by epistemic accounts of quantum states.

The Rule Perspective locates the chief difference between NQMCs and quantum states in their different \textit{linguistic roles} (descriptive vs. non-descriptive), not in a contrast between distinct realms of objects they are about. On the one hand, quantum states can legitimately (and with much theoretical gain) be assigned to objects of arbitrarily large ``macroscopic'' size, for instance by the help of the many-particle methods of quantum statistical mechanics. Their assignment is not confined to the objects of any putative ``micro-level''. On the other hand, even the most ``microscopic'' quantum objects elementary particle physicists have discovered (quarks, leptons, gauge bosons may be named) can very well be described in terms of NQMCs---with the caveat that not all NQMCs are licensed at all times---just as much as the largest objects of cosmology. There is even an interplay between the application of NQMCs and the assignment of quantum states to one and the same quantum system: Agents assign quantum states on the basis of their epistemic relations to the systems---that is, on the basis of what NQMCs they know to be true---, and the ``licensing'' of further NQMCs depends on the features of the quantum states they assign. The idea of a system to which \textit{only} quantum states and no NQMCs can be applied makes no sense on this view. The distinction between quantum states and non-quantum claims corresponds to a difference in linguistic roles, not to a difference between different types of objects referred to.

To conclude, epistemic accounts of quantum states are not committed to an ontological micro/macro divide. The Rule Perspective, in particular, does in no way suggest any two-level ontology where a ``classical'' macro-level contrasts with an ``unspeakable'' quantum micro-level. Furthermore, as I shall argue in what follows, there is no particular problem for non-ontic accounts of quantum states to account for reductive explanation of macro-properties in terms of the behaviour of micro-objects.

\subsection{Explanation without ontic quantum states}
Quantum theory is perhaps the theory with the greatest explanatory success in all the history of science. The final objection against epistemic accounts of quantum states to be discussed here is that non-descriptive readings of quantum states (not involving hidden variables) are \textit{in general} unable to account for quantum theory's incontestable explanatory force. Timpson raises this charge against quantum Bayesianism by (rhetorically) asking ``if quantum mechanics is not to be construed as a theory which involves ascribing properties to micro-objects along with laws describing how they behave, can we account for [its] explanatory strength?''\footnote{See \cite{Timpson} p.\ 600.} According to Timpson, the quantum Bayesian interpretation of quantum states as ``states of belief'' entails that all that can possibly be accounted for by means of quantum theoretical reasoning involving quantum states are agents' beliefs and expectations, not physical phenomena themselves. As an example, Timpson considers the explanation provided by quantum many-particle theory as to why some solid materials (sodium, in his example) conduct electricity well, whereas others don't, and remarks that ``[u]ltimately we are just not interested in agents' expectation that matter structured like sodium would conduct; we are interested in \textit{why in fact it does so}.''\footnote{Ibid. p.\ 600.} Quantum theory, as Timpson emphasises, helps us explain the behaviour of physical systems only because its vocabulary refers to these systems themselves, not to the scientists and their expectations and beliefs about them. Any interpretation of quantum theory that attempts to account for its explanatory force must respect this.

There is a natural response by means of which quantum Bayesians may attempt to address this challenge. It starts with the observation that on the quantum Bayesian account of the quantum theoretical formalism as a ``manual [...] to help make wiser decisions''\footnote{See \cite{Fuchs2010} p.\ 8.} the theory does not say anything about what agents actually \textit{do} believe, but rather about what they \textit{should} believe in which circumstances to maximise their predictive success. On this reading, quantum theory does not so much \textit{describe} our expectations and beliefs as it rather \textit{prescribes} what we should expect and believe under which conditions. Quantum Bayesians may therefore suggest that any quantum theoretical reasoning which leads to the expectation that some physical phenomenon or regularity will occur is a candidate quantum theoretical explanation of precisely that phenomenon or regularity, contrary to what Timpson suggests.

This suggestion leads directly into the troubled waters of the long-standing debate in the philosophy of science concerning the relation between explanation and prediction, which is too involved to be discussed here thoroughly. To keep matters simple, however, we can proceed by assuming that the quantum Bayesian may simply regard it as a \textit{ramification} of her view that to account for why some phenomenon or regularity is to be expected when correctly applying quantum theory (and in that sense predicting it quantum theoretically) is simply what it \textit{means} to explain that phenomenon or regularity quantum theoretically. In this vein, Richard Healey bases his pragmatist account of ``how quantum theory helps us explain''\footnote{See the title of \cite{Healey1}.} on the idea that ``[t]o use a theory to explain a regularity involves showing the regularity is just what one \textit{should expect} in the circumstances, if one accepts that theory.''\footnote{See \cite{Healey1} p.\ 6, emphasis mine.} To apply this idea to the example proposed by Timpson, consider an agent who competently applies quantum many-particle theory and arrives at the expectation that matter having the structure and composition of sodium should conduct electricity well. The quantum theoretical reasoning used is not \textit{about} the agent's expectations and beliefs, but it certainly functions as a guide for the agent when forming them. Why not count it as a candidate quantum theoretical explanation of the phenomenon or regularity at issue? It seems that Quantum Bayesians may well claim that they are able to account for quantum theory's explanatory force since they do not construe the theory as describing what agents actually \textit{do} believe and expect, but, instead, what they \textit{should} believe and expect.

Unfortunately, however, this response is not open to the quantum Bayesians, and the main reason for why not is again their rejection of the notion of a state assignment being performed correctly. On their view, it does not make any sense to ask whether some line of reasoning that involves the assignment of quantum states is correct or incorrect, and this means that quantum theory does not actually have the prescriptive bite which the line of defence just considered attributes to it. In other words, quantum Bayesianism does not have the conceptual resources to distinguish between failed and successful explanations in quantum theory. Arguably, this means that it cannot account for quantum theoretical explanation in general, and this vindicates Timpson's charge.

The Rule Perspective has much better resources to account for the notion of a quantum theoretical explanation, as it acknowledges from the start that the question of correctness does apply to lines of quantum theoretical reasoning which may lead to expectations about physical phenomena and regularities. In accordance with Healey's account as formulated in \cite{Healey1}, it emphasises the importance of NQMCs to describe both the phenomena and regularities for which quantum theoretical explanations are sought (the explananda) and the assumptions and background conditions on which the suggested explanations are based (the explanantia). Regarding the example of explaining the experimentally determined conductivities of solid materials, the explanantia include descriptions of the internal composition and structure of the materials together with information about the external conditions the materials are exposed to. Given these ``known facts'' about the systems at issue, the rules of state assignment dictate which quantum states to assign (though approximations may have to be made in practice due to the enormous difficulties involved in computing these states exactly). The Born Rule then permits to derive the probabilities for the possible values of observables from these states or, what amounts to the same, it permits to compute expectation values for these observables.\footnote{Quantum Bayesianism concurs that the Born Rule has the normative role of prescribing which probabilities to assign on the basis of which quantum states (see \cite{Fuchs2010} p.\ 8). It denies, however, that the theory has any normative force concerning which states are to be assigned on the basis of which evidence. This denial creates the problems discussed over and over in the present paper.} One such expectation value is the quantum theoretically predicted value for the conductivity. If it is in agreement with what is measured experimentally, the latter may count as explained by quantum theory---or so proponents of the Rule Perspective might suggest.

However, the idea that a quantum theoretical explanation is what makes an explanandum phenomenon or regularity \textit{expected} can be nothing more than a first approximation to a more accurate picture. In the remaining paragraphs of this section I discuss two respects in which it needs refinement and argue that the Rule Perspective has the conceptual resources to refine it as necessary. The two aspects correspond to two criticisms of Hempel's IS- (\textit{inductive statistical}) model of probabilistic explanation made by Peter Railton\footnote{See \cite{Railton} p.\ 211 f.} and David Lewis. According to Lewis, Hempel's model has the following two ``unwelcome consequences'':
\begin{quote}
(1) An improbable event cannot be explained at all. (2) One requirement for a correct explanation [...] is relative to our state of knowledge; so that or ignorance can make correct an explanation that would be incorrect if we knew more. Surely what's true is rather that ignorance can make an explanation seem to be correct when really it is not. (\cite{Lewis_expl} p.\ 232.)
\end{quote}
The claim that a quantum theoretical explanation is any line of quantum theoretical reasoning that makes an explanandum phenomenon or regularity expected has the same ``unwelcome consequences'' as Hempel's IS-model.

Let us briefly address the second ``unwelcome consequence'' first. It seems plausible that a line of reasoning which makes some phenomenon or regularity expected, but which allows for the possibility that this may no longer be so once more is known about the system at issue does not qualify as a candidate quantum theoretical explanation. The Rule Perspective does not have any problems to formulate this requirement: the rules of state assignment prescribe what quantum state assignment should be made under which epistemic conditions, and they apply whether or not these conditions are actually realised. In particular, one can meaningfully ask what expectations an agents \textit{should have} if she had whatever \textit{additional} information that is not excluded by the current available data. So, the Rule Perspective can require that a successful explanation need not only make the explanandum expected but that it should, moreover, ensure that the explanandum remain expected if the agent's epistemic situation were to improve in whatever conceivable manner.

Lewis' and Railton's first point concerns the explainability of improbable events such as decays of radioactive nuclei with long half-lifes. For these nuclei, considered individually, their decay is not to be expected on the time scales of a typical human life. There are quantum theoretical computations of the probabilities of these decays, which start from assumptions concerning the structure and interactions between the constituents of these nuclei. One may feel that these computations are explanatorily relevant as regards even the most rare nuclear decays and conclude that this disqualifies any perspective on quantum theoretical explanation that requires that a high probability be conferred on the explanandum.\footnote{This is the main motivation for Railton's deductive-nomological account of probabilistic explanation (see \cite{Railton}), which is based, however, on a propensity interpretation of probabilities that seems not straightforwardly combinable with the account of quantum probabilities given by the Rule Perspective.}

To respond to this challenge, the proponent of the Rule Perspective may deny that single event such as individual radioactive decays are in all cases candidate recipients of quantum theoretical explanations at all. Probabilistic theories such as quantum theory, one may argue, can in general explain only \textit{patterns} or \textit{regularities} among events.\footnote{See \cite{Woodward} for general considerations on explanation in probabilistic theories, which support this point of view.} As Healey notes, ``the phenomena that physicists are primarily concerned to explain are not particular individual happenings but general regularities in the properties or behavior of physical systems of a certain kind''\footnote{See \cite{Healey1} p.\ 2.}, which is why he confines his account of ``how quantum theory helps us explain'' to the explanation of regularities. This perspective still allows us to regard individual events, no matter how low their (nonzero) probabilities, as in an interesting sense \textit{accounted for} by quantum theory. In particular, this concerns individual radioactive decays. Given the distribution of potential energy in a nucleus, they would not occur if classical mechanics were the correct theory of nuclear systems, but they are allowed in quantum theory due to a phenomenon which is widely known as ``quantum tunnelling'': the penetration of a system (in this case, for instance, an alpha particle) through a potential barrier which is larger than the system's total energy. The misleading impression that genuine \textit{explanations} of individual radioactive decays are given by quantum theory, one may argue, arises from the fact that quantum theory does not \textit{exclude} them, whereas classical physics would.

There are \textit{some} events, however, for which quantum theory does seem to deliver single-case explanations, and these explanations should be accounted for by the Rule Perspective. In addition to the \textit{unsolicited} radioactive decays considered so far there are also so-called \textit{stimulated} radioactive decays, i.e. cases where radioactive decay is \textit{induced} by absorption of a neutron by a nucleus.\footnote{Nuclear fission is the most important technological application of stimulated decay.} It seems natural to regard quantum theory as accounting for how stimulated decays are \textit{caused} by incoming neutrons and as thereby delivering \textit{causal explanations} for these decays. Causal explanation, however, is widely regarded as ``\textit{ontic}'' explanation in the sense that the correctness of such an explanation does not depend on the epistemic conditions of those who give or receive it. It may therefore seem that an ontic account of quantum states is needed to accommodate for these causal explanations by quantum theory.

Fortunately, however, there are accounts of causation which conceive of ``cause'' and ``effect'' in such a way that causal explanations using quantum states are intelligible on an epistemic account of quantum states. An account of causation for which this is the case is the \textit{agency theory} of causation developed by Peter Menzies and Huw Price. It spells out causation in terms of ``agent probabilities'', where, as explained by Menzies and Price, ``the agent probability that one should ascribe to $B$ conditional on $A$ (which we symbolize as `$P_A(B)$') is the probability that $B$ would hold were one to choose to realize $A$.''\footnote{See \cite{Menzies} p.\ 190.} The core statement of the agency theory of causation is that $A$ causes $B$ if and only if $P_A(B)>P(B)$ for the ``agent probability'' $P_A(B)$. As Price makes clear in a more recent statement of his position, one important aspect of the ``agentiveness'' of agent probabilities is ``that they are assessed from the \textit{agent's} distinctive epistemic perspective.''\footnote{See \cite{Price} p.\ 494. The emphasis is Price's.} This strongly suggests that quantum probabilities, as conceived from the point of view of the Rule Perspective, can play the role of agent probabilities in the sense of the agency theory of causation. In particular, the probability which an agent ascribes on the basis of her quantum theoretical computations to the stimulated decay of a radioactive nucleus can function precisely as an agent probability in the sense of Menzies and Price. Even if that probability remains low (yet enhanced as compared to a situation where no stimulation via neutron absorption occurs), the decay counts as ``caused'' by the incoming neutron on the agency theory of causation combined with the Rule Perspective's view of quantum probabilities. This suggests that individual events such as stimulated decays can be regarded as \textit{explained} by quantum theory on a causal account of explanation that goes neatly together with the Rule Perspective.

``Explanation'' is a vast and multifaceted notion, and we should not expect to receive an account of what constitutes a quantum theoretical explanation in a single criterion.\footnote{A further desideratum for quantum theoretical explanations to be satisfying is perhaps that they help us attain a more \textit{unified} understanding of what is to be explained. Spelling out what this means, however, seems no more difficult on an epistemic than an ontic account of quantum states, which is why I neglect this issue for the purposes of the present investigation.} However, as the considerations just presented suggest, the Rule Perspective does not have any systematic difficulty to account for the notion of a quantum theoretical explanation. Furthermore, as argued in the previous subsection, it is not committed to an ontological micro/macro divide and is therefore no less well compatible with reductionist accounts of quantum theoretical explanation than the most-discussed ontic accounts of quantum states. As this shows, epistemic account of quantum states need not go hand in hand with the explanatory anti-reductionism advocated by the quantum Bayesians.\footnote{See \cite{Fuchs2010} p.\ 21 fn.\ 31 for a quantum Bayesian endorsement of explanatory anti-reductionism and \cite{Timpson} p.\ 592 and p.\ 600 for helpful remarks on the role and importance of anti-reductionism in quantum Bayesianism.} Reductive explanation can have its place in non-ontic as much as in ontic accounts of quantum states.

\section{Concluding Remark}
The aim of this paper has been to make it plausible that a therapeutic approach to quantum theory can be coherently articulated and defended in form of the Rule Perspective. Before closing this paper, I would like to acknowledge that the therapeutic accounts based on the epistemic conception of quantum states are not the only interpretations which attempt to make sense of quantum theory without introducing additional technical vocabulary such as hidden variables or explicit dynamics of collapse. Similar things can be said of the Everett interpretation, whose proponents regard it as the most striking virtue of their interpretation that it does not add any surplus theoretical elements to the original formalism of the theory. In the words of David Wallace:
\begin{quote}
If I were to pick one theme as central to the tangled development of the Everett interpretation of quantum mechanics, it would probably be: \textit{the formalism is to be left alone}. What distinguished Everett's original paper both from the Dirac-von Neumann collapse-of-the-wavefunction orthodoxy and from contemporary rivals such as the de Broglie-Bohm theory was its insistence that unitary quantum mechanics need not be supplemented in any way (whether by hidden variables, by new dynamical processes, or whatever). (See \cite{Wallace07} p.\ 311.)
\end{quote}
The idea of ``leaving the formalism alone'', however, is much less straightforward than Wallace suggests. Taken by itself, the formalism is just an uninterpreted piece of mathematics. Its extra-mathematical significance derives only from its application by competent physicists in quantum theoretical practice. Wallace's case for the Everett interpretation as the single outstanding take on quantum theory which does not ``supplement'' the formalism ``in any way'' is valid only if one presupposes that the role of quantum states is to describe the physical facts. Making this assumption is a natural first step when interpreting the theory, and to outline its implications is both highly important and rewarding, as the work of Everettians impressively demonstrates. For those impressed by the problems encountered by Everettians when trying to ``leave alone'' the formalism while giving it a descriptive reading\footnote{See, for instance, the criticisms of the Everett interpretation brought forward by Kent, Maudlin, and Price in \cite{saunders}.} a natural next step is to look for alternatives which also ``leave alone'' the formalism but refrain from construing quantum states as descriptive. The aim of the present paper has been to substantiate the claim that such alternatives exist and that at least one of them should be taken seriously.

\section*{Acknowledgements} I would like to thank Jeremy Butterfield for helpful comments on parts of an earlier draft of this article.

\end{document}